\documentclass[useAMS,usenatbib]{mn2e}
\usepackage{graphicx}

\title[Asymmetric Wolf-Rayet winds: implications for GRB afterglows.]{Asymmetric Wolf-Rayet winds: implications for GRB afterglows.}
\author[J.J. Eldridge]{J. J. Eldridge\thanks{E-mail: j.eldridge@qub.ac.uk}.\\
Astronomy Research Centre, School of Maths \& Physics, Queen's University Belfast,Belfast, BT7 1NN, Northern Ireland, UK\\}

\begin{document}

\date{}

\pagerange{\pageref{firstpage}--\pageref{lastpage}} \pubyear{2002}

\maketitle

\label{firstpage}

\begin{abstract}
Recent observations of Wolf-Rayet (WR) binaries WR151 and WR155 infer that their stellar winds are asymmetric. We show that such asymmetries can alter the stellar-wind bubble structure, bringing the wind-termination shock closer to the WR star. If the wind asymmetry is caused by rotation, the wind density and distance to the wind-termination shock are both decreased along the rotation axis by a factor of a few for the observed equator-to-pole wind density ratio of WR151. If this asymmetry lasts until core-collapse the time taken to reach the wind-termination shock by supernova ejecta or a gamma-ray burst jet is reduced. This leads to a distorted structure of the supernova ejecta and makes it more likely a constant density environment is inferred from gamma-ray burst afterglow observations.
\end{abstract}


\section{Introduction}

Wolf-Rayet (WR) stars are the most massive stars in the final stages of evolution and have lost (or in the process of losing the last remnants of) their hydrogen envelopes. Observationally they are very luminous, with broad emission lines indicating a fast and dense wind. WR stars are the preferred progenitors of type Ibc supernovae (SNe) due to the lack of a hydrogen envelope. They have also become the leading candidates to be the progenitors of Gamma-ray bursts (GRB). A GRB is thought to occur at core-collapse if a black hole forms and the core material surrounding the black hole possesses enough angular momentum to reside in an accretion disk. This disk can then feed material to the black hole resulting in a highly relativistic jet. If the progenitor is compact the jet will emerge at the surface to produce the prompt emission and later the afterglow. The supporting evidence is that GRBs occur in star forming regions of their host galaxies, some GRBs have associated type Ibc SNe and some afterglow lightcurves are consistent with the jet propagating in a free-wind environment expected around massive stars (for a recent review see \citet{review}). The evidence for stellar progenitors from absorption lines in afterglow spectra has weaken due to the recent findings of \citet{chen}.

However when the circumburst environment (CBE) density structure is inferred from afterglow observations some GRB are found to occur in constant density environments rather than a freely expanding wind where $\rho \propto 1/r^{2}$ \citep{grbp2,pana1,vm2}. Stellar-wind bubbles do have a region where the wind density structure is approximately a constant density. The expanding wind goes through a wind-termination shock (a hydraulic jump) when the ram pressure of the free-wind drops below the pressure of the stalled-wind region \citep{windbubbles1}. This stalled-wind region has a density profile that is roughly constant. Therefore if the free-wind region is small, the GRB jet can traverse it before the afterglow emission begins and a constant density environment would be inferred. Simulations show it is very difficult to shrink the free-wind region to a small enough radius for it to remain unobserved in the GRB afterglow \citep{vm2,egrb}. The required radius is roughly $10^{18}{\rm cm}$, a fraction of a parsec \citep{grbp2}. There are a number of possible physical processes that might shrink the free-wind region as outlined by \citet{vm2} but factors must be combined to produce a free-wind region with the required radius.

One possible process that may have an effect is asymmetry in WR star winds. The cause of asymmetric winds would typically be the result of stellar rotation or duplicity, however, as we discuss below there is some uncertainty in predicting how rotation affects the WR winds. Wind asymmetry can be estimated by performing polarimetry on stars as asymmetry induces strong polarisation \citep{polar}. Polarimetry surveys of WR stars have been performed and for most WR stars the polarisation is low with only 20 percent of the stars showing any noticeable intrinsic polarisation \citep{polarsurvey}. The observations indicate a mass-loss asymmetry between the highest and lowest wind density to be between a factor of 2 to 3. There is one important case of stronger polarisation from the close WR binary WR151 \citep{wr151}. In this system the light is polarised by 4 percent and modelling indicates that the equator-to-pole density ratio ($\alpha$) is 5.

\citet{wr151} suggest that the difference is due to rotation redistributing the mass-loss from the pole to the equatorial region as in the wind-compressed zone model of \citet{wcz}. If this is the case then there is a direct implication for the CBE of GRB progenitors which are thought to be rapidly rotating WR stars but the effect of stellar rotation on stellar winds is somewhat uncertain. The effect on B star winds has been well studied. \citet{bcwcd} and \citet{wcz} first suggested that inertial effects due to rotation reduce the polar wind density and increase the equatorial wind density. A more rigorous approach from \citet{owo1}, \citet{owo2} and \citet{pp00} suggest the converse is true when the distortion of the stellar surface by rotation and the radiative driving force are taken into account.

However only \citet{wcz} have considered WR winds. WR winds are different from winds of other hot stars as they are optically thick and hydrodynamic models of the winds are quite complex and computational intensive \citep{wcwinds}. The results for B star winds cannot be directly applied to WR winds because although they are both line driven the optical thickness may cause important differences and possibly result in the inertial terms dominating and causing the winds to be distorted as predicted by \citet{wcz}.

In this letter we use the observed asymmetry of WR151's wind to produce a model wind bubble for this system and study the effect which the asymmetry has on the bubble structure. However we do not consider what might cause the asymmetry. This avoids the uncertainty in the different models for wind distortion by rotation. We then perform a small parameter survey varying the initial interstellar medium (ISM) density and the size of the wind-density asymmetry and estimate the time it would take a GRB jet to reach the stalled-wind region.

\section{The current and future stellar wind bubble of WR151.}

\begin{figure*}
\includegraphics[angle=0, height=70mm]{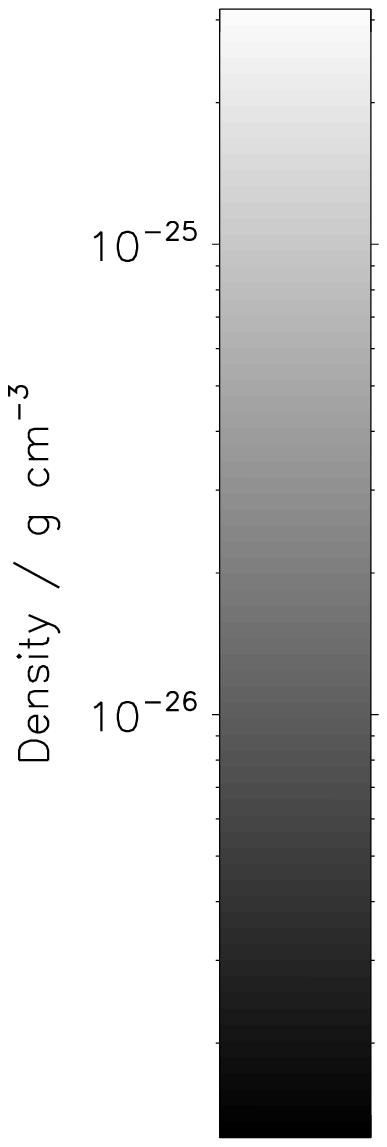}
\includegraphics[angle=0, height=75mm]{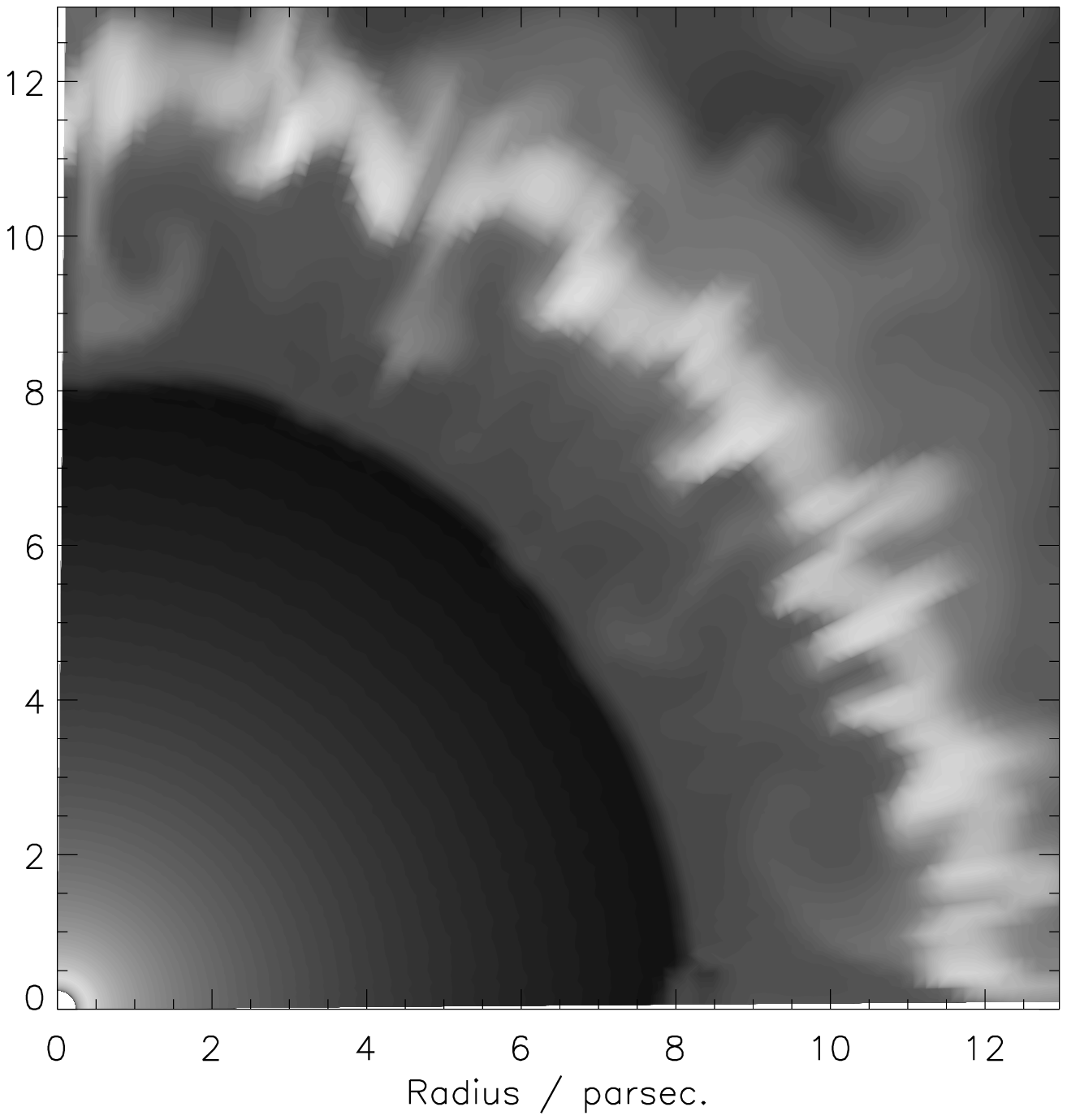}
\includegraphics[angle=0, height=75mm]{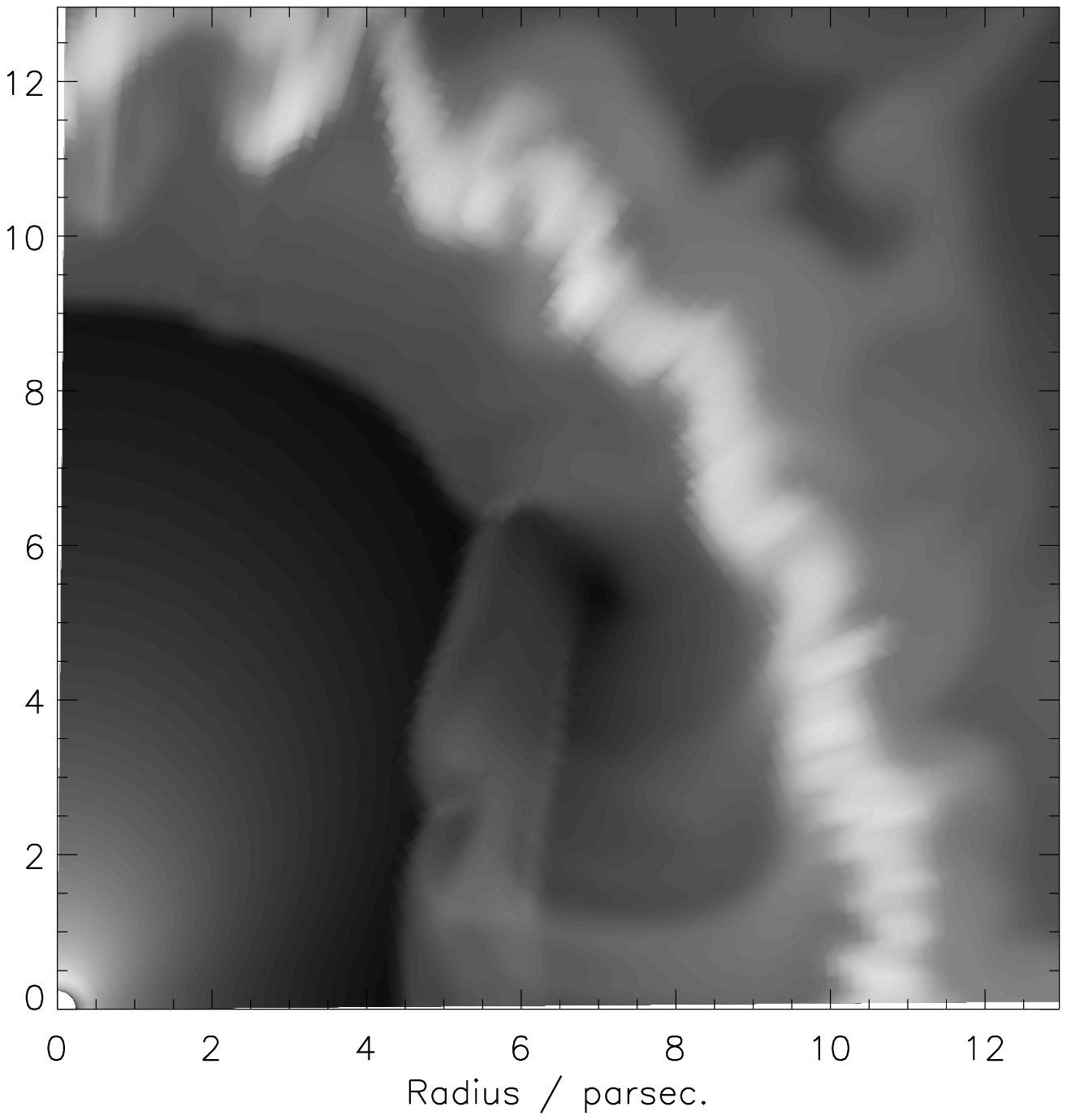}
\caption{Greyscale plots of the stellar-wind bubble structure within 12.5 parsecs of the WR star. The WR star had an initial mass of 60M$_{\odot}$ and is pictured here at an age of 3.8Myrs, the initial ISM density was $10 {\rm cm}^{-3}$. The left panel shows the case for a spherically symmetric wind while the right panel shows the case when the equator-to-pole wind density ratio, $\alpha= 5$. The polar direction is along the x-axis in both figures.}
\label{contournow}
\end{figure*}

First to model the stellar-wind bubble we must select a stellar model to use. In WR151 the current mass of the WR star is estimated to be 23M$_{\odot}$. To obtain the correct mass from our single star model we adopt a solar metallicity ($Z=0.02$) star with an initial mass of 60M$_{\odot}$ that has a long WN phase. We therefore assume that WR151 was initially a wide binary where the 60M$_{\odot}$ star became a red supergiant, filling it's Roche Lobe and growing to a radius greater than the binary orbit so that common envelope evolution occurs. During this the orbit shrank to its current size, the hydrogen envelope was lost and the helium core became a WN star. For the secondary, we assume that the primary's hydrogen envelope was lost rapidly and the secondary accreted very little material from the primary. Therefore we assume that the initial mass of the secondary was similar to its current mass of 30M$_{\odot}$. When we compare the wind strengths we find the primary wind dominates so we ignore the secondary's winds when modelling the wind bubble.

Our stellar-wind bubble model is made using the same code as described in \citet{egrb}. We use the mass-loss rates and wind velocities from the solar metallicity 60M$_{\odot}$model with an initial ISM density of $n_{0}=10{\rm cm}^{-3}$. We turn on the wind asymmetry when the star has lost its hydrogen envelope and assume that $\dot{M}(\theta) \propto 1+\beta (\sin \theta)^{2}$ where beta is $\beta=\alpha-1$. We normalise the mass-loss rate so that the total mass-loss from the star is unchanged from the symmetric case. Also we assume that the wind speed is unchanged. We select the stellar-wind bubble model with an age of 3.8 Myrs as the mass-loss rate and wind velocity are broadly in agreement with the observed parameters of WR151.

Figure \ref{contournow} shows a greyscale plot of the wind bubble density structure with and without the asymmetric mass-loss. The dense shell in the stalled-wind region is the RSG wind that has been swept up by the faster WR wind. The difference between the symmetric and asymmetric cases is clear with the distance to the wind termination shock in the polar direction reduced by nearly a factor of two. This is due to the reduction in wind density in the polar direction and therefore a reduction in the ram pressure compared to the stalled-wind ambient pressure which remains constant.

\begin{figure}
\includegraphics[angle=0, width=84mm]{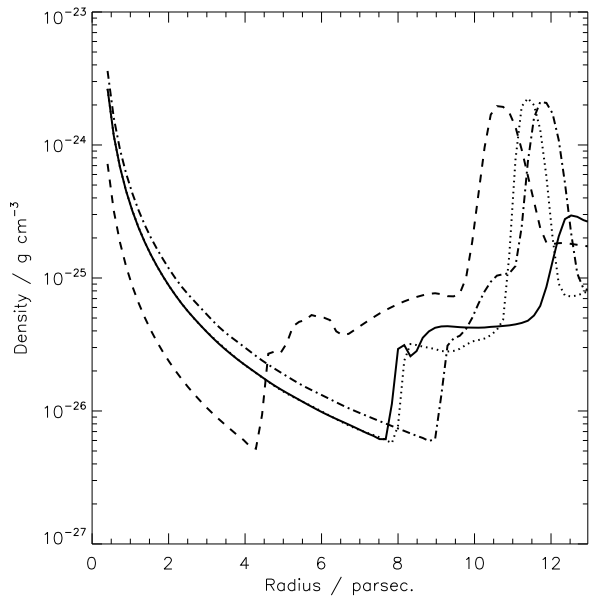}
\caption{Density profiles through the simulations shown in Figure \ref{contournow}. The solid and dotted lines are through the spherically-symmetric wind while the dashed and dash-dotted lines are through the asymmetric simulation along the pole and equator directions respectively.}
\label{profilenow}
\end{figure}

Figure \ref{profilenow} compares density profiles through this environment. In this plot it is easier to compare the change in the position of the wind termination shock in the polar and equatorial directions. It is important to note that while the polar distance is decreased the equatorial distance to the wind termination shock is increased because the slightly higher density of the wind leads to a greater ram pressure. Also in these plots the asymmetry we have introduced has only just been switched on within the simulation and therefore the stalled wind region is similar between the two simulations.

\begin{figure*}
\includegraphics[angle=0, height=70mm]{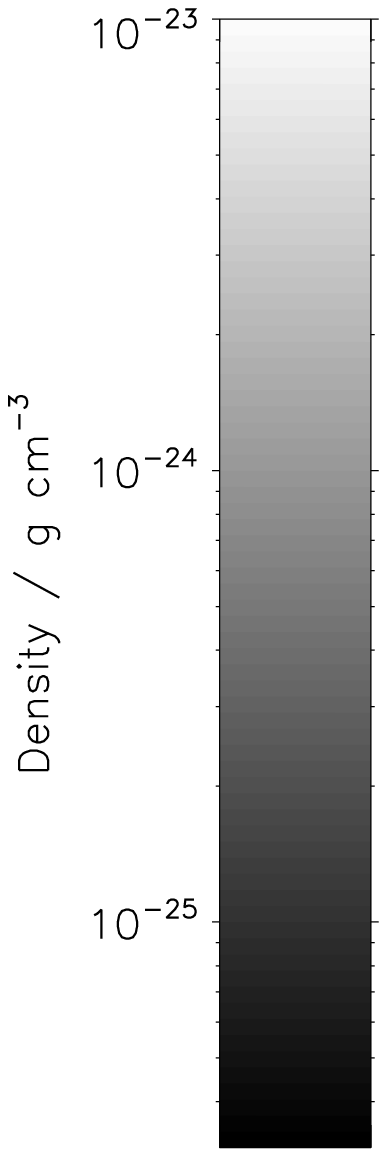}
\includegraphics[angle=0, height=75mm]{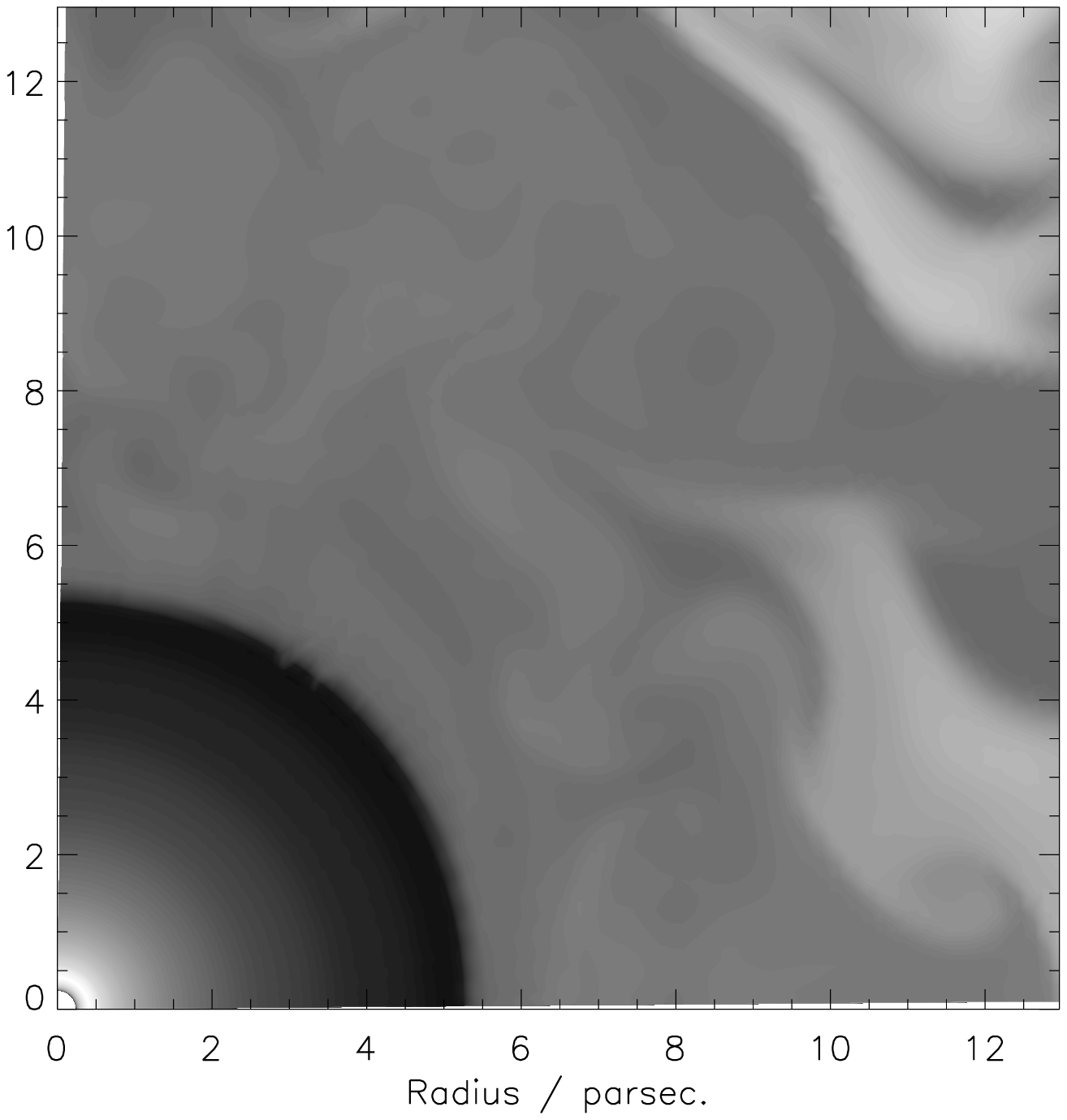}
\includegraphics[angle=0, height=75mm]{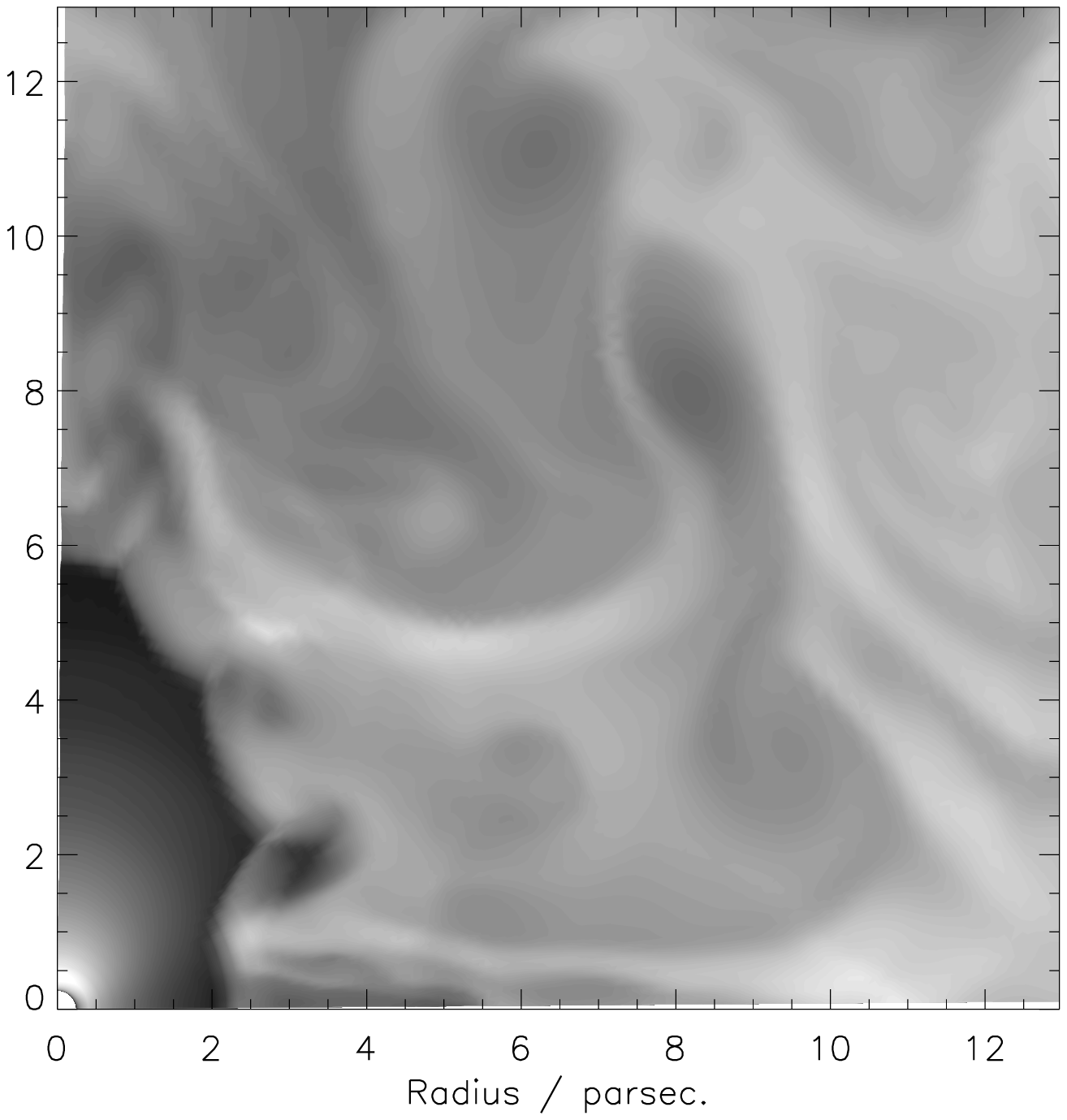}
\caption{Similar to Figure \ref{contournow} but 340,000 years later, a few years before core-collapse. The left panel shows the case for a spherically symmetric wind while the right panel shows the case when the equator-to-pole wind density ratio is 5. The polar direction is along the x-axis in both figures.}
\label{finalcontour}
\end{figure*}

\begin{figure}
\includegraphics[angle=0, width=84mm]{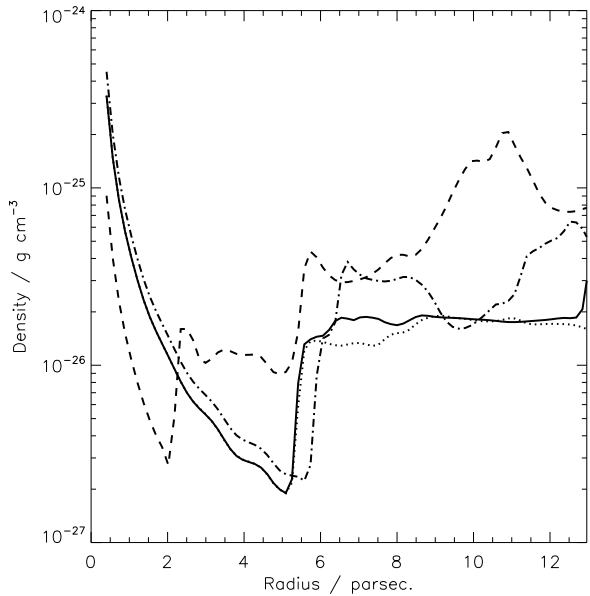}
\caption{Density profiles through the simulations shown in Figure \ref{finalcontour}. The solid and dotted lines are through the spherically-symmetric wind simulation while the dashed and dash-dotted lines are through the asymmetric simulation along the pole and equator directions respectively.}
\label{finalprofile}
\end{figure}

If we continue the simulations for another 0.34 Myrs up until the end of the stellar models (about a year before core-collapse) and we assume the asymmetry remains the distortion increases as the stalled-wind region is also altered. This is important as the supernova or gamma-ray burst that may occur at that time will propagate through this environment. The resulting wind-bubble structures are shown in Figures \ref{finalcontour} and \ref{finalprofile}. Both with and without the asymmetry the stalled-wind region has some denser swirls. These are due to instabilities in the wind \citep{garcia2} that have grown to produce great density variations as found by \citet{egrb}. We find with the asymmetric wind acting for a long time the magnitude and number of these swirls are greater. Overall we find the distance to the wind-termination shock, $R_{\rm SW}$ is decreased in the polar direction by a factor of approximately 3.

If a SN was to occur in such a medium the ejecta would reach the wind-termination shock in the polar direction sooner than along the equatorial region and therefore decelerate more quickly by accreting more material. The SN remnant would be distorted to an oblate shape rather than being spherically symmetric. If a GRB was to occur in such a structure the afterglow jet would reach the stalled-wind region in a shorter time and therefore the afterglow would appear to be traversing a constant density environment. This is in comparison to the symmetric wind bubble case where a freely expanding wind would be inferred. 

There is another reason why the $R_{\rm SW}$ is more uneven in Figure \ref{finalcontour}. \citet{egrb} found that the position of the wind-termination shock is not alway directly related to the initial ISM density as the analytic relation of \citet{windbubbles1}. This is because if the wind speed and wind density of the source star are varying on the same timescale as the time it takes for the wind to reach the wind termination shock then the boundary will move. As that time now varies between the pole and equator so too does the magnitude of this timescale effect.

\section{A parameter search.}

We have performed a parameter search varying the wind asymmetry, $\alpha$, and the initial ISM density, $n_{0}$, around the star to investigate the resulting change in the polar distance to the wind-termination shock. In table \ref{distances} we show the distance in parsecs to the wind termination shock and the wind density along the polar direction. Clearly the effect of increasing the wind asymmetry is that both $R_{SW}$ and $A_{*}$ decrease. The effect of increasing the initial ISM density decreases $R_{SW}$, as expected. There is one exception when at the highest ISM density, $R_{SW}$ increases. This is due to the wind-terminating shock becoming highly distorted so the time for the wind to travel from the star to the shock varies from pole to equator. Therefore if the wind is varying on a similar timescale the wind-termination shock will respond to any change on a different timescales.

With these distances and wind densities it is possible to estimate how long it would take for a GRB jet to traverse the distance to the beginning of the stalled-wind at $R_{\rm SW}$. Rather than perform a detailed afterglow calculation we use a simple model of isotropic relativistic ejecta with $E_{\rm iso}=10^{53} {\rm ergs}$ and $\Gamma_{\rm initial}=100$. As the ejecta expands it sweeps up material decreasing it's velocity. Therefore shorter distances and lower wind-densities both shorten the time for the jet to reach $R_{\rm SW}$. We note that we calculate the observer time that includes the effect of time dilation by viewing along the expanding ejecta direction of motion. However it does not include cosmological time-dilation therefore these times are a lower limit and will be longer for higher redshifts.

Table \ref{distances} show that the asymmetries greatly reduce the time for the jet to reach $R_{\rm SW}$. However a high density ISM is still required to reduce the size of the wind-bubbles. The time of around an hour is short enough that the early X-ray afterglow of GRBs may show some sign of occurring in a free-wind region while later emission may show a constant density profile. This agrees with \citet{pana2} where for the first few hours of a sample of GRB afterglows a free-wind CBE is preferred while after a few days a constant density CBE become more common \citep{pana1}. Therefore we may see the transition from free-wind to stalled-wind regions in some GRB afterglows without any bump in the lightcurve when the jet encounters the wind termination shock. It is important to note that our time estimate is approximate, if for example the Lorentz factor or the kinetic energy of the ejecta is higher then the times can be reduced further. Although our results demonstrate that even moderate asymmetries in the wind reduce the times by an order of magnitude.

\begin{table}
\caption{Position of the wind-termination shock from the WR stars with different initial ISM densities and different equator-to-pole density ratios during the WR phase. Also listed is the wind density in the polar direction for the different density ratios, $A_{*}=(\dot{M}/10^{-5}M_{\odot} \,{\rm yr^{-1}})/(v_{\rm wind} / 1000 {\rm km \, s^{-1}})$. Also show is the time taken for ejecta to reach the wind-termination shock $R_{\rm SW}$ for a simple isotropic model with $E_{\rm iso}=10^{53} {\rm ergs}$ and $\Gamma_{\rm initial}=100$.}
\label{distances}
\begin{tabular}{@{}ccccccccc@{}}
\hline
&$n_{0}/{\rm cm^{-3}}=$&   $10^{3}$ & &  $10^{2}$ & & $10^{1}$ &  \\
\hline
   & &  $R_{SW}$  & $t_{\rm SW}$  &  $R_{SW}$  & $t_{\rm SW}$  &  $R_{SW}$  & $t_{\rm SW}$    \\
$\alpha$ & $A_{*}$  &  /pc & /hrs &  /pc & /hrs &  /pc & /hrs  \\

\hline
1 & 0.66 & 0.76 & 24.4 & 2.67 & 794  & 5.25 & 5831 \\
2 & 0.33 & 0.47 &  7.2 & 2.02 & 356  & 3.80 & 2227 \\
3 & 0.22 & 0.37 &  4.1 & 1.89 & 295  & 3.31 & 1485 \\
5 & 0.13 & 0.31 &  2.7 & 1.41 & 130  & 2.18 & 443  \\
7 & 0.09 & 0.25 &  1.6 & 1.08 & 62.2 & 2.18 & 443  \\
10& 0.07 & 0.31 &  2.7 & 1.08 & 62.2 & 1.70 & 219  \\
15& 0.04 & 0.22 &  1.3 & 0.70 & 19.7 & 1.05 & 57.6 \\
20& 0.03 & 0.20 &  1.1 & 0.66 & 16.9 & 1.05 & 57.6 \\
\hline

\end{tabular}
\end{table} 

\section{Discussion and Conclusions.}

While the asymmetry in WR winds of WR151 and other WR stars could be due to rotation, duplicity or some other factor, it is the asymmetry itself that has a large effect on the stellar-wind bubble structure. It may also affect the evolution of the star since angular momentum loss will depend on how much mass is lost at equatorial latitudes \citep{mmgrb}. It will effect the evolution of a SN or GRB if it occurs within such a distorted wind-bubble. Spherically symmetric SN ejecta will be forced into an oblate form. For a GRB if we assume that the observed asymmetry is caused by rotation and that the direction of the lowest wind density is the rotation axis then the GRB jet will propagate in that direction. Therefore the more asymmetric the WR wind the more quickly it will encounter the wind-termination shock taking only a few hours. Furthermore because the wind density is also reduced the jet will also be decelerated less as it will sweep up less material. 

These factors together indicate that the afterglow jet is more likely to reach the constant density environment quickly so the free-wind environment does not affect the afterglow. This is suggested by the times listed in Table \ref{distances}. Combined with the other possible mechanisms discussed by \citet{vm2} to reduce the wind-termination shock distance we should not be surprised that many GRB afterglows apparently occur in constant density environments rather than free-wind environments. Also there is some evidence that stellar wind mass-loss rates are lower than currently used in stellar models with stars losing most of their mass in explosive outbursts \citep{lbv}. If mass-loss rates and therefore wind densities are reduced again a constant density environment becomes more likely.

We must also consider that GRB progenitors may require very rapid rotation. If this is the case then the models of \citet{wcz} may not apply as pointed out by \citet{owo1} and \citet{owo2}. If this is the case the reverse of our argument will be true and the wind termination shock distance along the polar direction will be much greater and the wind much denser. It is difficult to know exactly which model is correct, not because the models of the wind acceleration are limited but rather because the stellar evolution models are limited by being only 1D models. In the case of rapid rotation, the stellar structure will become distorted and 1D models give some indication of what might occur but they cannot fully model this inherently 3D problem.

By using the observed results on WR winds of what we assume is rotation (or duplicity) then we bypass this uncertainty. One remaining problem is that the WR stars observed to have asymmetries to date are solar metallicity WR stars in the Milky Way and it is though that GRB progenitors are lower metallicity WR stars. Therefore a study into the polarisation and asymmetry of lower metallicity WR stars in the LMC and SMC would be of great interest (e.g. Vink, in prep.).

In conclusion some WR stars are observed to have asymmetric winds and this has a large effect on the position of the wind-termination shock. The reduction of the wind density in one direction can lead to the wind-termination shock moving closer to the star. If a GRB jet was to travel in this direction then with the lower wind density and shorter distance, a constant density CBE is more likely to be inferred from the observations of the afterglow.

\section*{Acknowledgements}
This work, conducted as part of the award ``Understanding the lives of massive stars from birth to supernovae'' made under European Science Foundation EURYI Awards scheme. JJE would also like to thank Allard-Jan van Marle and Norbert Langer for useful discussion, Seppo Mattila for the careful proof reading this letter and the referee for their comments that have improved this letter.

\bsp

\label{lastpage}

\end{document}